\documentclass[aps,prl,twocolumn,10pt,superscriptaddress,floatfix,longbibliography]{revtex4-2}
\usepackage{amsmath,amssymb,bm,graphicx}
\usepackage{xcolor}
\usepackage[colorlinks=true,citecolor=blue,urlcolor=black,linkcolor=black]{hyperref}
\hypersetup{
  pdftitle={Fractional Charge at Dislocations from Band-Inversion Cancellations},
  pdfauthor={Vladimir Juri\v{c}i\'c and Rodrigo Soto-Garrido}
}
\newcommand{\bk}{\mathbf{k}}
\newcommand{\bb}{\mathbf{b}}
\newcommand{\bp}{\mathbf{p}}

\newcommand{\rev}[1]{{\color{black}#1}}
\newcommand{\new}[1]{{\color{black}#1}}
\newcommand{\chg}[1]{{\color{black}#1}}
\newcommand{\cy}[1]{{\color{black}#1}}
\newcommand{\rd}[1]{{\color{black}#1}}
\newcommand{\bl}[1]{{\color{black}#1}}

\expandafter\def\csname citediff@Pletyukhov2020Boundary\endcsname{}
\expandafter\def\csname citediff@Yoshida2023Polarization\endcsname{}
\expandafter\def\csname citediff@Vaidya2024\endcsname{}
\expandafter\def\csname citediff@RamamurthyHughes2015\endcsname{}
\begin{document}
\title{{Fractional Charge at Dislocations from Band-Inversion Cancellation}}
\author{Vladimir Juri\v{c}i\'c}
\email{vladimir.juricic@usm.cl}
\affiliation{Departamento de F\'isica, Universidad T\'ecnica Federico Santa Mar\'ia, Casilla 110, Valpara\'iso, Chile}
\author{Rodrigo Soto-Garrido}
\email{rodsoto@uc.cl}
\affiliation{Facultad de F\'isica, Pontificia Universidad Cat\'olica de Chile, Vicu\~na Mackenna 4860, Santiago, Chile}
\date{\today}
\begin{abstract}
Band inversions with opposite Chern numbers cancel in the Hall response and yield counterpropagating boundary modes that can gap out. However, a bulk probe of the topology that survives such a cancellation is lacking. We here show that dislocations in inversion-symmetric two-band crystals with cancelling band inversions bind a half-integer charge, which reveals the quantized bulk polarization of the bond-centered obstructed atomic limit realized by such crystals. To demonstrate this, we continuously deform a broad class of two-band lattice models into decoupled chains, without closing the bulk gap or breaking the inversion symmetry, where the dislocation charge follows from a count of occupied bonds. Notably, the fractional part of this charge is insensitive to local perturbations of the dislocation core, and it therefore persists after the in-gap core state is removed, as we show analytically in the chain limit and numerically away from it. In the Weyl-semimetal counterpart, the response to a magnetic flux separates this crystalline charge from the charge set by the Chern number of each momentum slice. We propose that a half-integer core charge that persists after the in-gap peak disappears from the local spectrum can be detected in artificial electronic and optical lattices, with metamaterials providing complementary access to the core modes.
\end{abstract}

\maketitle

\emph{Introduction.} Band inversions are the basic ingredient of topological band structures: in a two-band insulator a single inversion typically yields a quantized Hall conductance with a chiral edge mode~\cite{HasanKane2010,QiZhang2011}, and in three dimensions Weyl nodes with an anomalous Hall response~\cite{ArmitageMeleVishwanath2018}. This direct correspondence becomes more subtle when the boundary modes can be gapped out, as in higher-order topological insulators~\cite{Benalcazar2017Science,Schindler2018,Song2017Rotation,Geier2018,Khalaf2018,Calugaru2019,Trifunovic2019}. Another route opens when two band inversions contribute opposite Chern numbers: their Hall responses cancel, while the counterpropagating boundary modes hybridize on selected terminations~\cite{Langbehn2017,Khalaf2021}, suppressing the low-energy spectral weight near certain corners and hinges in anti--higher-order topological insulators and Weyl semimetals, respectively~\cite{MiaoTI2026,MiaoWeyl2026}. Nevertheless, such a crystal need not be topologically trivial, since the occupied charge can be centered between the lattice sites, yielding a bulk polarization quantized by the inversion symmetry~\cite{HughesProdanBernevig2011,Zak1989} and an obstructed atomic limit~\cite{Bradlyn2017TQC}. This calls for a bulk probe of the polarization that relies neither on transport nor on the boundary spectrum.

\begin{figure}[t!]
\includegraphics[width=\columnwidth]{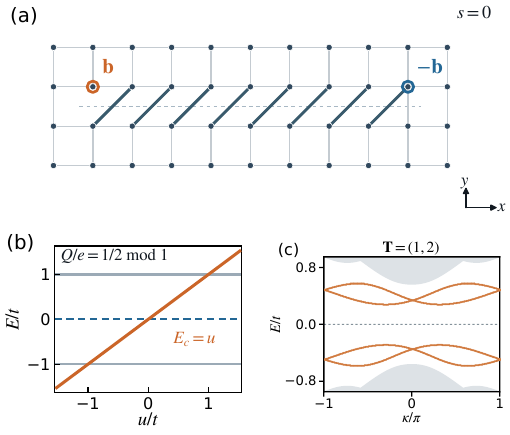}
\caption{Dislocation geometry and spectral control.
(a) The dislocation pair in the two-dimensional model
($s=0$). Between the defect cores, upward bonds across the dashed line
reconnect one cell to the right. Core 1 (orange) has
$\bb=(1,0)$ and core 2 (blue) has $-\bb$. Figure~\ref{fig:charge}(c)
shows the same geometry in the chain limit.
(b) At $s=1$ [Eq.~\eqref{eq:homotopy}], the potential $u$ shifts
the core-1 level to $E_c=u$ (orange); the core-2 level stays at zero
(blue dashed). For $|u|>t$ the core-1 level lies beyond the flat
bands at $\pm t$ (gray), with $Q/e=1/2$ modulo one
[Eq.~\eqref{eq:singlecharge}; End Matter, Sec.~\ref{app:charge}].
(c) The clean ribbon with edge translation $\mathbf T=(1,2)$ in the $k_z=0.2\pi$ slice of the Weyl model [Eq.~\eqref{eq:weyl}], which belongs to the sector with $\bp=(1/2,0)$, has a gap of about
$0.56t$ at half filling. Orange curves are edge states and
gray shading marks bulk bands. Here
$m_W=-1$, $B_x=1.5$, $B_y=0.75$, and $B_z=t=1$
(Sec.~VII~B and Figs.~S5 and S7 of the Supplemental
Material~\cite{SM}).}
\label{fig:translation}
\end{figure}

Dislocations offer a route to detect band topology directly in the bulk. A circuit around a dislocation core features a missing or an excess lattice translation by the Burgers vector $\bb$ relative to the same circuit in the perfect crystal. Bloch states near a band inversion at momentum $\mathbf K$ thereby acquire the phase $e^{i\mathbf K\cdot\bb}$, which underlies the $\mathbf K\cdot\bb=\pi$ (modulo $2\pi$) rule for in-gap dislocation modes, and has established dislocations as probes of translationally active, crystalline, and higher-order topological phases~\cite{Ran2009,Juricic2012,TeoKane2010,Slager2014,Roy2021,Queiroz2019,Malave2024}. Besides such modes, a dislocation can also bind charge: the same lattice mismatch leaves the bulk polarization uncompensated at the core, so that the number of electrons near the core acquires a fractional part fixed by the polarization~\cite{vanMiertOrtix2018,Schindler2022,Zhou2026}. In contrast to the modes, which local spectroscopy resolves in energy, this charge counts all occupied states and can therefore persist without an in-gap state~\cite{Schindler2022,Zhou2026}. Nevertheless, the dislocation response after the cancellation of band inversions has not been addressed. In particular, with the Hall response absent and the edge gapped, it is not clear which of the two, the core mode or the core charge, survives, and what it reveals about the bulk.

\begin{figure}[!t]
\includegraphics[width=\columnwidth]{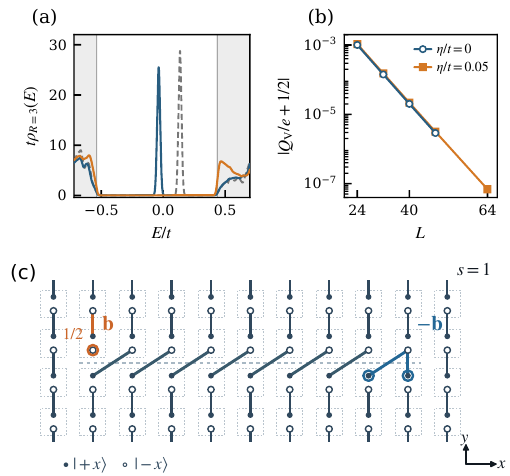}
\caption{Half-integer charge without an in-gap state at core 1. The model [Eq.~\eqref{eq:2d}] has $m=B=t=1$ and $J=0.9$.
Periodic $L\times L$ samples use the bond reconnection in
Fig.~\ref{fig:translation}(a).
(a) Local spectra with scalar hopping $\eta\cos k_x\sigma_0$ at
$\eta=0.05t$. The gray dashed curve is core 1 at
$u=t$ ($L=40$); orange and blue are cores 1 and 2,
respectively, at $u=12t$ ($L=64$). Shading marks energies outside
the bulk gap, approximately $(-0.54,0.44)t$. Each eigenstate is
weighted by its probability within three lattice spacings of the
core; Gaussian broadening $0.012t$ is used for display.
(b) At $u=12t$, the charge of core 1, $Q_{\rm V}/e$, approaches
$-1/2$ for $\eta=0$ and $0.05t$, without integer shifts, consistent
with Eq.~\eqref{eq:singlecharge} modulo one. The charge sums the
excess over one electron per cell in the region nearer core 1, with half weight for equidistant cells. The vertical axis
shows the absolute deviation from $-1/2$.
(c) The same lattice geometry as Fig.~\ref{fig:translation}(a) at
$s=1$ and $\eta=0$ [Eq.~\eqref{eq:homotopy}]. Each dashed box
is one cell. Solid and open circles denote $|+x\rangle$ and
$|-x\rangle$ at the same physical origin, drawn apart for clarity.
Each ordinary dimer shares one electron between two cells
[Eq.~\eqref{eq:dimer}]. At core 1, the remaining bond
(orange) supplies half an electron to the cell, and the orange
ring marks the unpaired orbital at energy $u$. At core 2,
two bonds meet at a common orbital. The three-orbital cluster has
energies $0,\pm\sqrt{2}t$; its zero mode has opposite amplitudes on
the two blue-ringed orbitals and vanishes on the common orbital.}
\label{fig:charge}
\end{figure}

\begin{figure}[!t]
\includegraphics[width=\columnwidth]{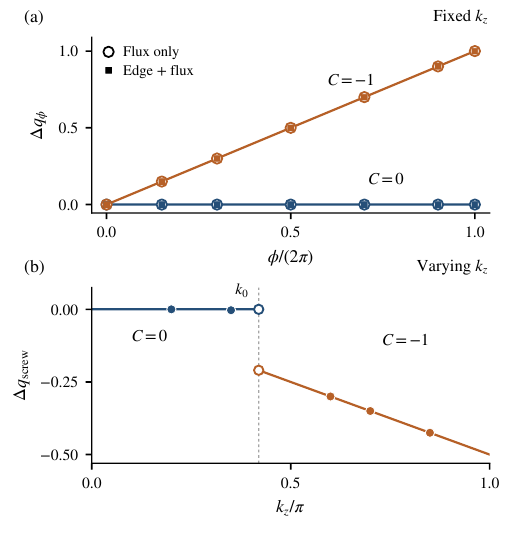}
\caption{\rd{Charge induced by applied flux and by the screw component of a
dislocation in the Weyl model [Eq.~\eqref{eq:weyl}]. Solid lines
show Eq.~\eqref{eq:flux}. Data use all occupied states in periodic
$32\times32$ slices, with occupations recalculated at each point
and charge branches unwrapped by integer shifts.
(a) At fixed $k_z$, flux is added in the dislocation-free lattice
(circles) or at an edge-dislocation core (squares).
We plot $\Delta q_\phi=[Q(\phi)-Q(0)]/e$, subtracting from each geometry its own zero-flux charge, which includes the charge already bound to the edge dislocation. Both geometries give zero at $k_z=0.2\pi$ (blue,
$C=0$) and $\phi/(2\pi)$ at $k_z=0.7\pi$ (orange, $C=-1$)
[Eq.~\eqref{eq:slices}], within numerical accuracy.
(b) The screw component of $\bb=(0,-1,-1)$ supplies the phase
$\phi=-k_z$ [Eq.~\eqref{eq:holonomy}]. The plotted
$\Delta q_{\rm screw}$ subtracts the charge at the edge-only core
$\bb=(0,-1,0)$ from that at the mixed core, at the same $k_z$ and
in units of $e$. The predicted difference is zero between the
Weyl nodes and $-k_z/(2\pi)$ outside them. Open markers show
the two limits at the node (dashed line at $k_0/\pi$).
This scan varies both the phase and the slice. End
Matter, Sec.~\ref{app:flux} gives the independent spectral-flow
counts.}}
\label{fig:flux}
\end{figure}

We here show that, for a pair of cancelling band inversions in a two-band crystal, the main dislocation response is a half-integer charge bound to the core, fixed by the bulk polarization that survives the cancellation, while the in-gap core mode need not survive. To this end, we deform the minimal two-band model with cancelling inversions into decoupled polarized
chains without closing the bulk gap or breaking the inversion symmetry. The deformation identifies the bulk phase as a bond-centered obstructed atomic limit, in which a dislocation that terminates a chain removes one bond’s half-electron contribution to the end cell~[Figs.~\ref{fig:translation}(a) and~\ref{fig:charge}(c)]. The half-electron offset survives even as a local potential
shifts the core level out of the gap, both in the chain limit
[Fig.~\ref{fig:translation}(b)] and in the full
two-dimensional crystal [Figs.~\ref{fig:charge}(a) and~\ref{fig:charge}(b)]. Finally, in the Weyl counterpart, inserting flux with and without a dislocation separates this charge from the Hall response of each momentum slice (Fig.~\ref{fig:flux}).

\emph{Minimal model.} As a starting point, we consider the minimal two-band lattice model featuring two cancelling band inversions~\textcolor{blue}{\cite{MiaoTI2026,MiaoWeyl2026}},
\begin{align}
 h_J&=t(\sin k_x\sigma_x+\sin k_y\sigma_y)+M_J(\bk)\sigma_z,
 \label{eq:2d}\\
 M_J(\bk)&=m-4B+2(B-J)\cos k_x+2(B+J)\cos k_y.\nonumber
\end{align}
Here, $t$ is the interorbital hopping amplitude, the parameters $m$, $B$, and $J$ set the momentum-dependent mass $M_J(\bk)$, with $J$ as the anisotropy between its $x$ and $y$ hopping terms, $\bk$ is the crystal momentum, the Pauli matrices $\sigma_{x,y,z}$ act in the orbital space, and we set the lattice constant to unity. Both orbitals are centered on the same lattice site, and one band is occupied (half filling). Crucially, the model is inversion symmetric, $Ph_J(\bk)P^{-1}=h_J(-\bk)$ with $P=\sigma_z$, so that its topology is encoded in the signs of the mass at the four inversion-invariant momenta. For $t,B>0$, $0<m<8B$, and $J>|m-4B|/4$, the masses $M_J(K)$ at $\Gamma=(0,0)$, $X=(\pi,0)$, $Y=(0,\pi)$, and $M=(\pi,\pi)$ have signs $(+,+,-,-)$, respectively. Relative to the all-negative reference, the $\Gamma$ and $X$ inversions contribute $-1$ and $+1$ to the Chern number, which therefore vanishes, $C=0$, together with the Hall conductance. Notice that for $J=0$ and $m\neq4B$ the masses at $X$ and $Y$ coincide, the number of inverted momenta is odd, and the model describes a Chern insulator. The anisotropy $J$ splits the masses at $X$ and $Y$, and thereby yields the pair of cancelling inversions.

The emergence of a fractional dislocation charge in such a phase can be understood from general principles. For a gapped Chern-zero crystal, let $\bp$ denote the total electronic polarization in lattice units relative to the fixed neutral reference. The inversion symmetry quantizes each component to $0$ or $1/2$ modulo one. A dislocation displaces the lattice by the Burgers vector $\bb$ along any circuit enclosing its core, and the polarization transverse to $\bb$ is then left uncompensated at the core, in the same way as at the edge of a crystal. The Burgers vector $\bb$ thus selects the charge~\cite{vanMiertOrtix2018,KingSmithVanderbilt1993,Resta1994,Song2021}
\begin{equation}
 \frac{Q_{\bb}}{e}=b_xp_y-b_yp_x\pmod1,
 \label{eq:singlecharge}
\end{equation}
where $Q_{\bb}/e$ counts the excess electrons at the core relative to the neutral reference, and $e>0$. Therefore, once the Hall response is cancelled, the fractional dislocation charge is determined by $\bp$ alone. The polarization left by the two cancelling inversions, and the location of the charge at the core, both follow from the chain limit of the model, which we discuss next.

\emph{Chain limit.} To show that the cancelling inversions leave behind a polarized phase, we continuously deform the model in Eq.~\eqref{eq:2d} into decoupled chains through the interpolation
\begin{align}
 h_s&=(1-s)t\sin k_x\sigma_x+t\sin k_y\sigma_y
       +M_s(\bk)\sigma_z,\label{eq:homotopy}\\
 M_s(\bk)&=(1-s)M_J(\bk)+st\cos k_y,\nonumber
\end{align}
with $0\leq s\leq1$. For $s=0$ we recover the two-dimensional model, while for $s=1$ the dependence on $k_x$ disappears and the crystal decomposes into decoupled chains along $y$. Importantly, the bulk gap remains open along the entire path. A gap closing would require $\sin k_y=0$. The mass $M_J$ depends linearly on $\cos k_x$, so its signs at $k_x=0,\pi$ extend along each line: positive at $k_y=0$, negative at $k_y=\pi$. The chain mass $t\cos k_y$ has the same signs, so the interpolation never passes through zero on either line. The gap therefore stays open. The inversion symmetry and the orbital positions remain fixed, and therefore the polarization cannot change along the path. The entire sector with the mass signs $(+,+,-,-)$ is thus in the same phase as the decoupled chains.

In the chain limit ($s=1$) the model is exactly solvable, since the only hopping left in Eq.~\eqref{eq:homotopy} pairs the orbital $|\mathbf r,+x\rangle$ with $|\mathbf r+\hat y,-x\rangle$ [Eq.~\eqref{eq:dimer}; Fig.~\ref{fig:charge}(c)], where $|\pm x\rangle=(|1\rangle\pm|2\rangle)/\sqrt2$ and $|1\rangle$, $|2\rangle$ are the two orbitals in one cell. Each such dimer has energies $\pm t$, and its occupied state shares one electron equally between the two cells~\cite{SSH1979,Zak1989}. The occupied charge is therefore centered on the bonds along $y$, halfway between the lattice sites, which yields $\bp=(0,1/2)$ and identifies the phase as a bond-centered obstructed atomic limit~\cite{Bradlyn2017TQC}. For $J<-|m-4B|/4$, the same construction gives bonds along $x$ and $\bp=(1/2,0)$. The benchmark gap along the path is given in End Matter, Sec.~\ref{app:charge}. The same construction applies to two-band models with longer-range hopping, with the mass staying positive (negative) along the line $k_y=0$ ($k_y=\pi$), as shown in Sec.~I of the Supplemental Material (SM)~\cite{SM}.

\emph{Fractional dislocation charge.} In the chain limit the dislocation charge follows from counting the occupied bonds. A bulk cell receives half an electron from each of two bonds and is neutral at one electron per cell. Figure~\ref{fig:translation}(a) shows the dislocation pair in the two-dimensional lattice. Between the two cores, bonds crossing the cut reconnect one cell to the right. No cells are removed. At $s=1$ this shift leaves one $|-x\rangle$ orbital without a bond at the core with Burgers vector $\bb$ (core 1), where a chain of dimers effectively ends, and attaches two bonds to a single orbital at the core with $-\bb$ (core 2) [Fig.~\ref{fig:charge}(c)]. The cell of core 1 that holds the unpaired orbital thus has one bond instead of two. The remaining bond contributes half an electron and the unpaired orbital contributes its occupation $n_c=0$ or $1$. Thus the excess electron number is $Q/e=(1/2+n_c)-1=n_c-1/2$~\cite{JackiwRebbi1976,GoldstoneWilczek1981,vanMiertOrtix2017}. Notice that occupying the unpaired orbital changes the electron number by an integer; the missing bond fixes its fractional part.

This count is the chain realization of the polarization law in Eq.~\eqref{eq:singlecharge}. For chains along $y$, $\bp=(0,1/2)$ and odd $b_x$ terminates an odd number of chains, giving half-integer charge; for chains along $x$, $\bp=(1/2,0)$ and odd $b_y$ plays this role. Furthermore, since the bulk gap and the inversion symmetry are maintained for $s<1$, the same fractional charge is bound to the dislocation in the two-dimensional model, where the transverse hopping only spreads the excess density over the nearby cells. The count uses the total polarization: two identical copies contribute two half-electron offsets at the same core, giving integer charge (Sec.~III~B of the SM~\cite{SM}). By contrast, a coupled four-band model with total polarization $\bp=(1/2,1/2)$ binds half-integer charge for $\bb=(1,0)$ and $(0,1)$, and integer charge for $\bb=(1,\pm1)$ (Sec.~V of the SM~\cite{SM}).

\emph{Fractional charge without an in-gap state.} The bond count shows that the fractional charge and the in-gap state have different origins: the former is carried by the occupied bonds, while the latter is the unpaired orbital. Therefore, a local perturbation should be able to remove the state from the gap and leave the fractional charge intact. We first show this in the chain limit, which lets us move the core level without changing any occupied bond away from the cores. The onsite term $u(\sigma_0-\sigma_x)/2$ acts only on the unpaired $|-x\rangle$ orbital of core 1, shifting it to $E_c=u$ without changing its wave function [Fig.~\ref{fig:translation}(b)]. Here $\sigma_0$ is the $2\times2$ identity matrix. At half filling of $N$ cells, the $N-1$ bonding states are occupied, leaving one electron for the two core levels. This electron occupies the core-1 orbital for $u<0$ and the zero mode of core 2 for $u>0$, so that the charge of core 1 changes from $+e/2$ to $-e/2$, i.e., by an integer. For $|u|>t$, the core-1 level lies outside the gap between the flat bands at $\pm t$, and this core therefore carries half-integer charge without an in-gap state. The bond count at both cores and the charge balance for an isolated dislocation are given in End Matter, Sec.~\ref{app:charge}.

To further support this scenario beyond the solvable limit, we therefore numerically analyze the two-dimensional model with $m=B=t=1$ and $J=0.9$, in the presence of the same core potential. We also add scalar hopping $\eta\cos k_x\sigma_0$, which preserves the inversion symmetry and the bulk eigenvectors while breaking the two-band spectral constraint [Eq.~\eqref{eq:spectralconstraint}]. At $\eta=0.05t$ and $u=t$, core 1 has an in-gap peak near $0.14t$. At $u=12t$ the shifted level lies above the bulk bands, and the only remaining in-gap state is at core 2 [Fig.~\ref{fig:charge}(a)]. The charge near core 1 nevertheless approaches $-e/2$ with or without scalar hopping [Fig.~\ref{fig:charge}(b); Sec.~II~B and Fig.~S4 of the SM~\cite{SM}]. More generally, a local perturbation at an isolated core changes its integrated charge only by an integer multiple of $e$, as long as the chemical potential remains in the bulk gap (End Matter, Sec.~\ref{app:charge}).

The same polarization also fixes the fractional excess charge of an insulating edge~\cite{VanderbiltKingSmith1993Surface,Song2021,ZhangBarkeshli2025Boundary}. Whether a given edge is insulating, however, depends on further details, in particular on whether the lattice translation $\mathbf T$ along the edge allows the two counterpropagating edge modes to hybridize. For instance, for $\bp=(1/2,0)$ a clean ribbon with $\mathbf T=(1,2)$ is gapped at half filling [Fig.~\ref{fig:translation}(c)], whereas the edge crossings survive for $\mathbf T=(1,1)$. The boundary spectrum alone therefore does not determine the polarization; the roles of the edge translation, the spectral constraint of the two-band model, and the filling are discussed in Secs.~III~B and VII of the SM~\cite{SM}.

\emph{Weyl semimetal and flux response.} We now turn to the three-dimensional counterpart of the model, the anti--higher-order Weyl semimetal~\cite{MiaoWeyl2026}. For a straight dislocation line along $z$, the momentum $k_z$ is conserved and labels two-dimensional slices, each of the form of Eq.~\eqref{eq:2d}, which are gapped away from the Weyl nodes~\cite{Bulmash2015,SotoGarrido2020,Gioia2021,Hirsbrunner2024,HirsbrunnerGray2024}. The two nodes, located at $k_z=\pm k_0$, divide these slices into two families [Eq.~\eqref{eq:slices}]. Between the nodes, two band inversions, at $\Gamma$ and $Y$, cancel, the slice Chern number vanishes, and each slice is in the polarized phase discussed above, here with $\bp=(1/2,0)$. A dislocation with odd $b_y$ therefore binds half-integer charge in every such slice. Outside the nodes a single inversion remains, and the slices carry the Chern number $C=-1$, which underlies the anomalous Hall response of the Weyl semimetal~\cite{Wan2011,BurkovBalents2011,GoswamiRoyDasSarma2017}. A dislocation in the Weyl crystal can thus carry two kinds of charge: the crystalline charge fixed by the polarization, and the charge stemming from the Hall response of the Chern slices. The question is then how to distinguish them. 

Their response to a magnetic flux provides the answer. A flux $\phi$ threading a slice pumps the charge $-C(k_z)\phi/(2\pi)$, in units of $e$, to the flux core~\cite{Laughlin1981,Carey2017} [Eq.~\eqref{eq:flux}]. We insert $\phi$ first in a dislocation-free slice and then at the core of an edge dislocation, and in each geometry measure the charge relative to its own value at zero flux. The increment is identical in the two geometries [Fig.~\ref{fig:flux}(a)]: zero in the Chern-zero slice, where the dislocation still carries half charge, and $\phi/(2\pi)$ in the $C=-1$ slice. Therefore, the flux response measures only the Chern number of the slice, whereas the polarization is encoded in the charge present at zero flux.

A screw component of the Burgers vector supplies such a phase without applied flux. Encircling a line with $\bb=(\bb_\perp,b_z)$ translates the lattice by $b_z$ along the line, so that a state with momentum $k_z$ acquires the phase $k_zb_z$, which enters the slice in the same way as $\phi$ [Eq.~\eqref{eq:holonomy}]. Comparing the mixed dislocation $\bb=(0,-1,-1)$ with its edge-only counterpart $(0,-1,0)$ at the same $k_z$ then isolates the charge induced by the screw phase $\phi=-k_z$. It follows the same flux law: it vanishes between the nodes and equals $-k_z/(2\pi)$ outside them [Fig.~\ref{fig:flux}(b)]. Unlike a fixed-slice flux insertion, this scan varies both the phase and the bulk slice. The Weyl Hamiltonian, the conditions on its parameters, the symmetries that quantize the slice polarization, and the numerical validation are given in End Matter, Secs.~\ref{app:weyl} and~\ref{app:flux}, and in {Secs.~VII~A and IX--XI} of the SM~\cite{SM}.

\emph{Discussion and outlook.} To summarize, we have shown that a dislocation binds a half-integer charge in crystals where two band inversions cancel in the Hall response. The chain limit makes its origin transparent: a missing bond at a dislocation exposes the polarization that survives band-inversion cancellation. The core level can move out of the gap while the half-electron offset remains; in a Weyl crystal, the charge added by flux instead measures the slice Chern number. The argument relies on the bulk gap, the inversion symmetry, and the signs of the mass along the lines $k_y=0$ and $k_y=\pi$ of the Brillouin zone, rather than on the details of the model, and the same dislocation charge therefore arises in a broader class of two-band crystals with cancelling band inversions and longer-range hopping (Sec.~I of the SM~\cite{SM}).

Experimentally, these predictions can be tested by tuning the core potential while tracking both the local spectrum and the integrated excess density: the in-gap peak should disappear from the local spectrum, while the integrated core charge remains half-integer. Artificial electronic lattices and tunnelling microscopy resolve local spectra~\cite{Drost2017,Nayak2019}, and site-resolved imaging in fermionic optical lattices measures occupation~\cite{Nakajima2016,Parsons2015}. Magnetomechanical, microwave, photonic, and acoustic lattices provide complementary ways to vary defect and boundary couplings and track their modes~\cite{Grinberg2020,Peterson2021,Li2018,Lin2023,Liu2024Dislocation,Hu2026,Ye2022}.

Beyond band insulators, a similar cancellation of Chern numbers occurs in the interaction-driven anti-topological electron crystal predicted in twisted bilayer MoTe$_2$~\cite{Reddy2026AntiCrystal}, where the many-body polarization~\cite{Resta1994,Song2021,ZhangBarkeshli2025} can be used to test whether dislocations bind a fractional charge also in a correlated state. In non-Hermitian lattices, dislocations can produce skin accumulation and bound states~\cite{Panigrahi2022,SchindlerPrem2021,Bhargava2021,Zhong2026}; whether a physically specified occupation yields an analogous defect charge remains open. Similarly, in periodically driven (Floquet) topological insulators, where dislocations bind normal and anomalous dynamical modes~\cite{NagRoy2021}, whether a dislocation charge can be defined for the Floquet bands remains open.

\emph{\bl{Acknowledgments.}} We are grateful to Bitan Roy \bl{for a critical reading} of the manuscript. This work is supported by Fondecyt (Chile) Grants  No. 1230933 (V.J.) and 1241033 (R.S.-G.). 

\nocite{Vaidya2025Crystalline,You2016,Huang2019Torsion,Fukui2005,Pletyukhov2020Boundary,Yoshida2023Polarization,Vaidya2024,RamamurthyHughes2015}
\bibliography{references}

\clearpage
{\section*{End Matter}}
\appendix
\renewcommand{\appendixname}{}
\setcounter{secnumdepth}{1}
{\section{Chain limit, charge reference, and local reconstruction}
\label{app:charge}
At the endpoint of Eq.~\eqref{eq:homotopy}, the intercell hopping is
\begin{equation}
 T_y=\frac{t}{2}(\sigma_z-i\sigma_y)=t|+x\rangle\langle-x|.
 \label{eq:dimer}
\end{equation}
The two orbital combinations remain embedded at the same cell origin; their separated symbols in Fig.~\ref{fig:charge}(c) indicate orbital content. Each dimer has one occupied state with equal weight in its neighboring cells. This occupied-state boundary count is consistent with Berry-phase formulations of excess charge~\cite{vanMiertOrtix2017,RhimBehrendsBardarson2017}; its Volterra completion gives Eq.~\eqref{eq:singlecharge} in our counterclockwise Burgers-loop convention~\cite{vanMiertOrtix2018}. A charge window contains the localized density while staying far from the partner or outer boundary.

\cy{The masses at $\Gamma,X$ and $Y,M$ have the same signs as the endpoint mass $t\cos k_y$, which establishes the gap along Eq.~\eqref{eq:homotopy} for the stated sector. Section~I~A of the SM~\cite{SM} proves the general criterion and its extension to an even scalar dispersion at insulating filling.} For $m=B=t=1$, $J=0.9$, full-Brillouin-zone minimization yields the benchmark result
\begin{equation}
 g(s)=\sqrt{\frac{20+15s}{84-49s}}\geq\sqrt{5/21}.
 \label{eq:benchmarkgap}
\end{equation}
\cy{This closed expression is specific to this benchmark; the mass-sign criterion proves positivity for the broader class. The minimization is given in Sec.~I~B of the SM~\cite{SM}.}

{The endpoint calculation uses $L=16$.} With the bond reconnection of Fig.~\ref{fig:translation}(a), $N$ retained cells give $N-2$ ordinary dimers, a three-orbital cluster at core 2 with energies $0,\pm\sqrt2t$, and the unpaired orbital of core 1 [Fig.~\ref{fig:charge}(c)]. The core-2 zero mode is the antisymmetric combination of the cluster's two $|+x\rangle$ orbitals. Thus $N-1$ negative-energy bonding states leave one electron for the two core modes. Filling the instantaneous lowest $N$ states occupies the core-1 orbital for $u<0$ and the unchanged core-2 mode for $u>0$, with $Q_1+Q_2=0$. At $u=0$, the zero-mode degeneracy requires an occupation convention. These are canonical ground states at each potential, rather than a prescribed dynamical transfer. A single-core sample balances charge at a boundary or reservoir; its integer branch depends on the filling condition. Sec.~II of the SM~\cite{SM} derives the exact cluster count.}

For two isolated-core realizations differing by a finite-support Hermitian perturbation, with chemical potential in their common essential bulk gap, the localized Fermi-projector difference has integer relative trace~\cite{AvronSeilerSimon1994}. Local reconstruction changes the total charge of an isolated core by an integer multiple of $e$; a finite integration region may retain small density tails. Local crossings change the integer occupation branch. This argument fixes the fractional class only at the same crystal Hamiltonian far from the core, orbital embedding, charge reference, and Burgers vector; \new{adding an extended polarized chain changes the defect construction.} \new{Bulk crystalline symmetry quantizes polarization without requiring each local core reconstruction to preserve that symmetry.} \cy{The model $h_J$ obeys the two-band spectral constraint}
\cy{\begin{equation}
 \sigma_xh_J^*(\bk)\sigma_x=-h_J(-\bk),
 \label{eq:spectralconstraint}
\end{equation}}
\cy{which can pin levels or enforce zero-mode parity}~\cite{TeoKane2010,deJuan2014}, but allowed scalar core potentials break it while preserving the bulk polarization.

\section{Weyl model and mass sectors of the three-dimensional counterparts}
\label{app:weyl}

The Weyl Hamiltonian of Ref.~\cite{MiaoWeyl2026}, used in Figs.~\ref{fig:translation}(c) and~\ref{fig:flux}, is
\begin{equation}
 H_W=t(\sin k_x\sigma_x+\sin k_y\sigma_y)+M_W(\bk)\sigma_z,
 \label{eq:weyl}
\end{equation}
where $M_W(\bk)=m_W+\sum_{i=x,y,z}B_i\cos k_i$. Each slice at fixed $k_z$ is therefore a two-dimensional model of the form in Eq.~\eqref{eq:2d}, with $m-4B\to m_W+B_z\cos k_z$, $2(B-J)\to B_x$, and $2(B+J)\to B_y$, and its topology is again encoded in the signs of the mass at $\Gamma$, $X$, $Y$, and $M$. To follow these signs as functions of $k_z$, define
\begin{align}
 A_\Gamma&=m_W+B_x+B_y,& A_X&=m_W-B_x+B_y,\nonumber\\
 A_Y&=m_W+B_x-B_y,& A_M&=m_W-B_x-B_y.
\end{align}
With $t,B_z>0$, the condition
\begin{equation}
 |A_Y|<B_z<\min(A_\Gamma,-A_X,-A_M)
\end{equation}
keeps $\Gamma$ positive and $X,M$ negative while only $Y$ changes sign. Since zero energy requires $k_x,k_y\in\{0,\pi\}$, these inequalities enumerate all bulk closings. The mass at $Y=(0,\pi)$, $A_Y+B_z\cos k_z$, vanishes at $\cos k_0=-A_Y/B_z$, which gives two Weyl nodes at $(0,\pi,\pm k_0)$. For $m_W=-1$, $B_x=1.5$, $B_y=0.75$, and $B_z=t=1$, used throughout, $k_0=\arccos(1/4)$, and the slice Chern number is
\begin{equation}
 C(k_z)=
 \begin{cases}
 0,&|k_z|<k_0,\\
 -1,&|k_z|>k_0 .
 \end{cases}
 \label{eq:slices}
\end{equation}
Between the nodes the mass signs at $(\Gamma,X,Y,M)$ are $(+,-,+,-)$: the $\Gamma$ and $Y$ inversions cancel and leave $\bp=(1/2,0)$, the rotated version of the polarized phase of the main text. A dislocation with odd $b_y$ therefore carries half-integer charge there at zero flux. Outside the nodes the signs are $(+,-,-,-)$, only the $\Gamma$ inversion remains, and $C=-1$. Regarding the quantization of $p_x$ in the Chern-zero slices, notice that each slice of $H_W$ is inversion symmetric by itself, since $M_W$ depends on $k_z$ only through $\cos k_z$, and it additionally obeys the antiunitary symmetry in Eq.~\eqref{eq:Ay}; either of the two quantizes $p_x$. A perturbation that retains only the three-dimensional inversion, such as $\delta\sin k_z\sigma_y$, relates opposite momenta and does not quantize each slice (Secs.~IX and X of the SM~\cite{SM}). At equality, node creation or annihilation has zero longitudinal velocity and is not a simple Weyl point. Secs.~\rev{VI and VII} of the SM~\cite{SM} retain the full numerical phase benchmarks.

\section{Flux response, screw phase, and momentum-resolved occupation}
\label{app:flux}
To separate the Hall response from the crystalline charge, we hold $k_z$ fixed and insert a phase $\phi$, first in a dislocation-free slice and then at an edge-dislocation core. Subtracting the zero-flux charge in each geometry isolates the extra charge from the applied flux~\cite{Laughlin1981,Carey2017},
\begin{equation}
 q(k_z,\phi)-q(k_z,0)=-\frac{C(k_z)\phi}{2\pi}\pmod1,
 \label{eq:flux}
\end{equation}
where $q=Q/e$. The relation holds modulo one because a localized level crossing the chemical potential changes the charge by an integer. Over a full cycle, $\phi\to\phi+2\pi$, the net spectral flow at the chosen core is $-C(k_z)$. Equation~\eqref{eq:flux} contains no reference to the polarization: the charge already bound to an edge dislocation at $\phi=0$ drops out of the difference, which is why the two geometries in Fig.~\ref{fig:flux}(a) give the same increment.

\new{At each phase, the charge is computed by filling the same number of lowest-energy real-space states.} Spectral flow separately counts crossings through a reference energy; aligning charge branches does not impose adiabatically frozen occupations. At $L=32$, the four flux charge sweeps in Fig.~\ref{fig:flux}(a) satisfy Eq.~\eqref{eq:flux} \chg{with dimensionless charge errors below $7\times10^{-6}$.} The central mixed geometry has compensating crossings and zero net flow; \new{an outer slice has one signed crossing per core.}

A screw component supplies a phase without applied flux. A circuit around a line with Burgers vector $\bb=(\bb_\perp,b_z)$ gives
\begin{equation}
 U_{\bb}(k_z)=e^{ik_z b_z}T_{\bb_\perp},
 \label{eq:holonomy}
\end{equation}
where $T_{\bb_\perp}$ is the transverse lattice shift, the same as for an edge dislocation in the slice, and the factor $e^{ik_zb_z}$ arises from the translation by $b_z$ along the line. The core with $\bb=(0,-1,-1)$ thus has the phase $\phi=-k_z$, and subtracting the charge of the edge-only core $(0,-1,0)$ at the same $k_z$ isolates the screw contribution, which follows Eq.~\eqref{eq:flux} on either side of the nodes [Fig.~\ref{fig:flux}(b); \textcolor{blue}{Sec.~XI} of the SM~\cite{SM}]. For the selected member of $\pm(0,1,1)$, $\phi=-k_z$. At $k_z/\pi=0.6,0.7,0.85$, the differences from the edge-only core agree with $-k_z/(2\pi)$ within $5\times10^{-5}$. The central $k_z=0.35\pi$ point has \rev{$\Delta q_{\rm screw}\simeq-0.0030$} at $L=32$, {a finite-size deviation.} A fixed-slice flux index does not require a localized structural screw band at every outer momentum~\cite{Kodama2019}.

\bl{Within a fixed $k_z$ slice, $\mathcal A_y=P_xK$ combines
$x$ reflection $P_x$ with complex conjugation $K$ and imposes
\begin{equation}
 H_W(k_x,k_y;k_z)=H_W^*(k_x,-k_y;k_z).
 \label{eq:Ay}
\end{equation}}
\rev{The core-exchange realization of Eq.~\eqref{eq:Ay}, its allowed complex onsite terms, and the bulk-symmetry control are detailed in Sec.~\rev{IX} and Fig.~S3 of the SM~\cite{SM}.}

\rev{The momentum-resolved excess is}
\begin{align}
 \frac{Q_R(k_z)}{e}&=\sum_r w_R(r)\,\delta n_r(k_z),\nonumber\\
 \delta n_r(k_z)&=\int_{-\infty}^{\mu}d\omega\,A_r(k_z,\omega)-n_{{\rm ref},r}.
\end{align}
Here $A_r$ is the orbital-summed local spectral function, $w_R$ the core integration window, and $n_{{\rm ref},r}$ the matched neutral reference. \rev{A tunable two-dimensional realization with $k_z$ as an external parameter can access these slice quantities; a local probe of a three-dimensional line generally integrates over longitudinal momentum.}
\bl{For a straight line with conserved $k_z$, the charge per unit
length is obtained by integrating the unreduced slice charges at
a common chemical potential, including Chern slices and occupied
defect bands~\cite{Song2021,Wang2021Anomalies}. For the
inversion-preserving perturbation, the fractional charge shifts
cancel at opposite $k_z$ over a fixed symmetric interval of $C=0$
slices that remain gapped and Chern zero with and without the perturbation. This cancellation alone does not determine the integer
occupation branches or the physical line charge.}
Sec.~\rev{VIII} of the SM~\cite{SM} gives a finite-system check, and Sec.~\rev{IX}~A explains symmetric-interval cancellation.

\end{document}